\documentclass{webofc}
\usepackage[varg]{txfonts}   
\usepackage{float}

\usepackage{mathtools,amsmath,amssymb,amsfonts,dsfont,mathrsfs}
\usepackage{braket}
\usepackage{xparse}
\usepackage{breqn}
\usepackage[colorlinks=True,allcolors=blue]{hyperref}

\DeclareDocumentCommand{\Ag}{ s o }{ \IfBooleanTF{#1}
    { \IfValueTF{#2}{ \bm{\mathcal{A}}_{(#2)} }{ \bm{\mathcal{A}} } }
    { \IfValueTF{#2}{    {\mathcal{A}}_{(#2)} }{    {\mathcal{A}} } } }
\newcommand{\we}{\mathop{\scriptstyle\wedge}}

\NewDocumentCommand\MyAc{ m }{#1}
\DeclareDocumentCommand{\bt}{ t. t, t- s s m m m }{
  \RenewDocumentCommand\MyAc{ m }{##1}
  \IfBooleanT{#1}{\RenewDocumentCommand\MyAc{ m }{ \mathring{##1} } }
  \IfBooleanT{#2}{\RenewDocumentCommand\MyAc{ m }{ \tilde{##1} } }
  \IfBooleanT{#3}{\RenewDocumentCommand\MyAc{ m }{ \bar{##1} } }
  \IfBooleanTF{#4}
  { \IfBooleanTF{#5} { \hat{\MyAc{\mathcal{B}}}_{{#6}}{}^{\hat{#7}}{}_{\hat{#8}} }{ \hat{\MyAc{\mathcal{B}}}_{{#6}}{}^{{#7}}{}_{{#8}} } }
  { \MyAc{\mathcal{B}}_{{#6}}{}^{{#7}}{}_{{#8}} } }

\DeclareDocumentCommand{\ct}{ t. t, t- t' s s m m m }{
  \RenewDocumentCommand\MyAc{ m }{##1}
  \IfBooleanT{#1}{\RenewDocumentCommand\MyAc{ m }{ \mathring{##1} } }
  \IfBooleanT{#2}{\RenewDocumentCommand\MyAc{ m }{ \tilde{##1} } }
  \IfBooleanT{#3}{\RenewDocumentCommand\MyAc{ m }{ \bar{##1} } }
  \IfBooleanT{#4}{\RenewDocumentCommand\MyAc{ m }{ {##1}' } }
  \IfBooleanTF{#5}
  { \IfBooleanTF{#6} { \hat{\MyAc{\Gamma}}_{{#7}}{}^{\hat{#8}}{}_{\hat{#9}} }{ \hat{\MyAc{\Gamma}}_{{#7}}{}^{{#8}}{}_{{#9}} } }
  { \MyAc{\Gamma}_{{#7}}{}^{{#8}}{}_{{#9}} } }

\DeclareDocumentCommand{\ri}{ t. t, t- s s m m m }{
  \RenewDocumentCommand\MyAc{ m }{##1}
  \IfBooleanT{#1}{\RenewDocumentCommand\MyAc{ m }{ \mathring{##1} } }
  \IfBooleanT{#2}{\RenewDocumentCommand\MyAc{ m }{ \tilde{##1} } }
  \IfBooleanT{#3}{\RenewDocumentCommand\MyAc{ m }{ \bar{##1} } }
  \IfBooleanTF{#4}
  { \IfBooleanTF{#5} { \hat{\MyAc{\mathcal{R}}}_{{#6}}{}^{\hat{#7}}{}_{\hat{#8}} }{ \hat{\MyAc{\mathcal{R}}}_{{#6}}{}^{{#7}}{}_{{#8}} } }
  { \MyAc{\mathcal{R}}_{{#6}}{}^{{#7}}{}_{{#8}} } }

\newcommand*{\de}[1]{\mathop{\mathrm{d}#1}\nolimits}

\begin{document}
%

\fancyhead[L]{{\bf XXIII Chilean Physics Symposium} \\ \small Chile, 22-24 november  2022} 
\fancyhead[R]{\small Gravitation and Cosmology}
%
\title{Polynomial affine gravity in \(3+1\) dimensions}
%
%

\author{\firstname{Jos\'e}
  \lastname{Perdiguero}\inst{1}\fnsep\thanks{\email{jose.perdiguerog@gmail.com}}
  \and
  \firstname{Oscar}
  \lastname{Castillo-Felisola}\inst{1,2}\fnsep\thanks{\email{o.castillo.felisola@protonmail.com}}
}

\institute{Departamento de F\'isica, Universidad T\'{e}cnica Federico Santa Mar\'\i a, Casilla 110-V, Valpara\'iso, Chile
  \and
  Centro Cient\'ifico Tecnol\'ogico de Valpara\'iso, Casilla 110-V, Valpara\'\i so, Chile}

\abstract{The polynomial affine gravity is an alternative model of
  gravity whose fundamental field is the affine connection, and it is
  invariant under the complete group of diffeomorphisms. In \(3+1\)
  dimensions the field equations generalise those of Einstein--Hilbert,
   the coupling constants are dimensionless, the
  action has a finite numbers of term, and although the action does
  not involve a (fundamental) metric, some metric tensor fields might
  \emph{emerge} from the connection. Provided a cosmological ansatz,
  the properties of diverse cosmological models are discussed.}

\maketitle

\section{Introduction}
\label{sec:intro}

The relativistic equations for gravitational interactions were
introduced in 1915 by Einstein and Hilbert independently 
\cite{hilbert15_grund_physik,einstein15_grund_allgem_relat_und_anwen},
cementing the idea of geometrisation of gravity. The foundation of
General Relativity is Riemannian geometry, which is determined by the
metric tensor field. Such choice endows the metric with a double role:
(i) it is the object that serves to \emph{measure distances}, and (ii)
it is the mediator of gravitational interactions.

However, the development of the \emph{Erlangen Program} (initiated by
Klein \cite{klein08_compar_review_recen_resear_geomet} and concluded
by Cartan \cite{ivey03_cartan_begin}) yields the idea of
non-Riemannian spaces \cite{eisenhart27_non_rieman}, sometimes called
\emph{metric-affine} spaces
\cite{hehl95_metric_affin_gauge_theor_gravit}, inducing several
proposal generalising the model of Einstein--Hilbert. For example, Weyl
considered an affine connection which was not compatible with the
metric \cite{weyl18_reine_infin,weyl18_gravit_und_elekt}, Cartan
considered a non-symmetric connection
\cite{cartan23_sur_les_connex_affin_et,cartan24_sur_les_connex_affin_et,cartan25_sur_les_connex_affin_et},
Veblen proposed a projective version of General Relativity
\cite{veblen33_projec_gener_relat}, and Einstein and Eddington
proposed an affine formulation of General Relativity
\cite{einstein23_zur_affin_feldt,eddington23}, where the metric does
not play the role of mediating the interactions.

From those scenarios, the affine formulations of gravity are the more
general, since in principle the space do not require a (pre-existing)
metric structure. The polynomial affine model of gravity
\cite{castillo-felisola15_polyn_model_purel_affin_gravit,castillo-felisola18_einst_gravit_from_polyn_affin_model}
was built up as an attempt to obtain the less constrained model for
gravitational interactions.

In this letter we briefly discuss the properties of the cosmological
models found on the polynomial affine model of gravity complementing
the results from Ref.
\cite{castillo-felisola21_aspec_polyn_affin_model_gravit_three,castillo-felisola22_polyn_affin_model_gravit}.

\section{The model}
\label{sec:model}

As we mentioned, the polynomial affine model of gravity is a field
theory for the affine connection, \(\ct*{\mu}{\lambda}{\nu}\), and the most
general action can be built up using their irreducible components,
\(\ct{\mu}{\lambda}{\nu}\), \(\Ag_\mu\) and
\(\bt{\mu}{\lambda}{\nu}\), defined by the relation
\(\ct*{\mu}{\lambda}{\nu} = \ct{\mu}{\lambda}{\nu} + \bt{\mu}{\lambda}{\nu} + \delta^\lambda_{[\mu} \Ag_{\nu]}\),
and the volume element
\(\de{V}^{\alpha \beta \gamma \delta} = \mathrm{d}x^{\alpha} \we \mathrm{d}x^{\beta} \we
\mathrm{d}x^{\gamma} \we \mathrm{d}x^{\delta}\) (See
Ref.~\cite{castillo-felisola20_emerg_metric_geodes_analy_cosmol} for
details).

Up to boundary and topological terms, the action is written
as~\cite{castillo-felisola20_emerg_metric_geodes_analy_cosmol},
\begin{dmath*}[style={\small}]
  \label{eq:new-action}
  S  = \int \de{V}^{\alpha \beta \gamma \delta} \biggl[ B_1
  \ri{\mu\nu}{\mu}{\rho} \bt{\alpha}{\nu}{\beta}
  \bt{\gamma}{\rho}{\delta} + B_2 \ri{\alpha\beta}{\mu}{\rho}
  \bt{\gamma}{\nu}{\delta} \bt{\mu}{\rho}{\nu} + B_3
  \ri{\mu\nu}{\mu}{\alpha} \bt{\beta}{\nu}{\gamma} \Ag_\delta + B_4
  \ri{\alpha\beta}{\sigma}{\rho} \bt{\gamma}{\rho}{\delta} \Ag_\sigma
  + B_5 \ri{\alpha\beta}{\rho}{\rho}
  \bt{\gamma}{\sigma}{\delta} \Ag_\sigma + C_1
  \ri{\mu\alpha}{\mu}{\nu} \nabla_\beta \bt{\gamma}{\nu}{\delta} +
  C_2 \ri{\alpha\beta}{\rho}{\rho} \nabla_\sigma
  \bt{\gamma}{\sigma}{\delta} + D_1 \bt{\nu}{\mu}{\lambda}
  \bt{\mu}{\nu}{\alpha} \nabla_\beta \bt{\gamma}{\lambda}{\delta}
  + D_2 \bt{\alpha}{\mu}{\beta} \bt{\mu}{\lambda}{\nu}
  \nabla_\lambda \bt{\gamma}{\nu}{\delta} + D_3 \bt{\alpha}{\mu}{\nu}
  \bt{\beta}{\lambda}{\gamma} \nabla_\lambda \bt{\mu}{\nu}{\delta} +
  D_4 \bt{\alpha}{\lambda}{\beta} \bt{\gamma}{\sigma}{\delta}
  \nabla_\lambda \Ag_\sigma + D_5 \bt{\alpha}{\lambda}{\beta}
  \Ag_\sigma \nabla_\lambda \bt{\gamma}{\sigma}{\delta}
  + D_6 \bt{\alpha}{\lambda}{\beta} \Ag_\gamma \nabla_\lambda
  \Ag_\delta + D_7 \bt{\alpha}{\lambda}{\beta} \Ag_\lambda
  \nabla_\gamma \Ag_\delta + E_1 \nabla_\rho \bt{\alpha}{\rho}{\beta}
  \nabla_\sigma \bt{\gamma}{\sigma}{\delta} + E_2 \nabla_\rho
  \bt{\alpha}{\rho}{\beta} \nabla_\gamma \Ag_{\delta}
  + F_1 \bt{\alpha}{\mu}{\beta} \bt{\gamma}{\sigma}{\delta}
  \bt{\mu}{\lambda}{\rho} \bt{\sigma}{\rho}{\lambda} + F_2
  \bt{\alpha}{\mu}{\beta} \bt{\gamma}{\nu}{\lambda}
  \bt{\delta}{\lambda}{\rho} \bt{\mu}{\rho}{\nu} + F_3
  \bt{\nu}{\mu}{\lambda} \bt{\mu}{\nu}{\alpha}
  \bt{\beta}{\lambda}{\gamma} \Ag_\delta + F_4
  \bt{\alpha}{\mu}{\beta} \bt{\gamma}{\nu}{\delta} \Ag_\mu \Ag_\nu
  \biggr].
\end{dmath*}
One key feature of this purely affine formulation is that the number
of terms that can go to the action is limited, due to the restriction
of the \emph{index structural analysis} (for more details see Ref.
\cite{castillo-felisola18_einst_gravit_from_polyn_affin_model,castillo-felisola20_emerg_metric_geodes_analy_cosmol}).
This feature contrast strongly with other gravitational theories. For
example the Starobinsky model adds a correction
\(\alpha \mathcal{R}^{2}\) to the Einstein--Hilbert action, but there is no constraint
on the number of possible correction that can go the action. Whereas
in Polynomial Affine Gravity one has constraints coming from index
structural analysis. This is what we call the \emph{rigidity} of the
model.

The field equations for the model can be obtained by the standard
process of optimisation, but with the subtlety of preserving the
symmetries of the fields, e.g. the \(\bt{}{}{}\)-field is traceless.
Since the action depends on the first derivatives of the
fields, in principle there is no need of considering affine analogous
to the Gibbons--Hawking--York term.


In order to solve the field equations, it is customary to start from
an ansatz. For the sake of simplicity, we consider an affine
connection which is compatible with the cosmological principle, which
is parameterised as follows,
\begin{align*}
  {\Gamma_{t}{}^{t}{ }_{t}} & = f(t),
  &
    \Gamma_{i}{ }^{t}{ }_{j} & = g(t) S_{i j},
  &
    \Gamma_{t}{ }^{i}{ }_{j} & = h(t) \delta^{i}_{j},
  &
    \Gamma_{i}{ }^{j}{ }_{k} & = \gamma_{i}{ }^{j}{ }_{k},
  \\
  \mathcal{B}_{\theta}{ }^{r}{ }_{\varphi} & = \psi (t) r^2\sin\theta \sqrt{1 - \kappa r^2} ,
  &
    \mathcal{B}_{r}{}^{\theta}{}_{\varphi} & =\frac{\psi (t) \sin \theta}{\sqrt{1 - \kappa r^2}},
  &
    \mathcal{B}_{r}{}^{\varphi}{}_{\theta} & =\frac{\psi(t)}{ \sqrt{1-\kappa r^{2}} \sin \theta},
  &
    \mathcal{A}_{t} & = \eta(t),
\end{align*}
where \(S_{ij}\) and \(\gamma_{i}{}^{j}{}_{k}\) are the three dimensional
rank two symmetric tensor and connection compatible with the
cosmological symmetries. The function \(f\) can be set to zero by a
suitable re-parameterisation of \(t\). Given such choice, the function
\(f\) will not be present in the dynamic equations. 

Using the above ansatz, the cosmological field equations are given by
\begin{dgroup*}
  \begin{dmath*}
    0 = \left( B_3(2\kappa + gh + \dot{g}) + 2 B_4(gh - \dot{g}) + 2D_6 \eta g
      - 2F_3 \psi ^2 \right) \psi
  \end{dmath*}
  \begin{dmath*}
    0 = \left( \left( B_3 - 2B_4\right) \eta \psi - C_1 (2\eta \psi - \dot{\psi})
    \right) g
  \end{dmath*}
  \begin{dmath*}
    0 = \left(B_3 + 2 B_4 \right) \eta g \psi + 2 C_1 \left( \kappa \psi + 4gh \psi -
      g\dot{\psi} - \psi \dot{g} \right) + 2 \psi^3 \left( -D_1 + 2D_2 - D_3
    \right)
  \end{dmath*}
  \begin{dmath*}
    0 = B_3 \left( (h\psi - \dot{\psi}) \eta - \psi \dot{\eta} \right) - 2B_4 \left(
      (-h\psi - \dot{\psi})\eta - \psi \dot{\eta} \right) + C_1 \left( 4 h^2 \psi + 2 \psi
      \dot{h} - \ddot{\psi} \right) + D_6 \eta^2 \psi
  \end{dmath*}
  \begin{dmath*}
    0 = B_3(2\kappa + gh +\dot{g})\eta + 2 B_4 (gh - \dot{g})\eta + C_1 (2\kappa h + 4
    gh^2 + 2 g \dot{h} - \ddot{g}) + 6 h \psi^2 (-D_1 + 2D_2 - D_3) + D_6
    \eta^2 g - 6 F_3 \eta \psi^2
  \end{dmath*}
\end{dgroup*} 

\section{Final remarks}
\label{sec:remarks}

Despite the complexity of the field equations, solving the first
couple of equations provide four scenarios (given their
factorisation). However, the remaining equations from those scenarios
are still very loose, making very difficult to extract useful
information from them. This push us to consider other alternatives.

Another approach is to ask for compatibility between the connection
and the emergent metric tensor structure coming from either the Ricci
tensor \(\mathcal{R}_{\mu\nu}\) or the torsion-descent Pop{\l}aswki metric
\(\mathcal{P}_{\mu\nu} = \mathcal{A}_\mu \mathcal{A}_\nu + \mathcal{B}_{\mu}{}^{\lambda}{}_{\sigma}\mathcal{B}_{\nu}{}^{\sigma}{}_{\lambda}\)
\cite{poplawski14_affin_theor_gravit}. This additional condition
together with the field equations allows us to find analytically
solutions.

In the case of the Pop{\l}aswki, we have an \emph{affine scale factor}
given by the \(\psi(t)\) function, which serves as analogue of the scale
factor \(a(t)\) from the Friedman--Robertson--Walker metric. In that
sense, we are able to obtain solutions as (Anti) de Sitter, where the
parameter playing the role of cosmological constant comes from an
integration constant. This type of solution is considered to be an
eternal inflationary scenario, for \(\ddot{a} > 0\). It might be
argued that in this purely affine formulation of gravity we can have a
\emph{geometric inflation}, in the sense that our model does not
include matter.

Two of the most consolidated inflationary scenarios are the
power-law scale factor and the Starobinsky model. Those models
require the introduction of a self-interacting scalar field. A
relevant question is whether or not phenomenological viable scenarios
of \emph{geometric inflation} can be found in the context of
polynomial affine gravity, even in the absence of an energy-momentum
tensor. That question is part of the problems that drive our research.

\end{document}